# Contrasting Implications of the Frequentist and Bayesian Interpretations of Probability when Applied to Quantum Mechanics Theory

James L. Beck

*Engineering and Applied Science, California Institute of Technology, Pasadena, CA 91125*

An examination is made of the differing implications from applying the two mainstream interpretations of probability, frequentist and Bayesian, to QM (quantum mechanics) theory. As a test bed, the Bohm-EPR experiment is chosen and the second-order joint probability distribution from QM theory that describes the possible spin outcomes for two particles with coupled spins is examined. Several contrasting conclusions are made: (i) Under the frequentist interpretation where probability distributions are viewed as properties of inherently random processes, the QM spin distribution implies a widely-discussed non-locality because probabilistic conditioning on a spin measurement is viewed as corresponding to a causal influence. Under the Bayesian interpretation where probability distributions are viewed as a measure of the relative plausibility of each possible spin outcome, this conditioning is viewed as providing information relevant to the spin probabilities and the argument for non-locality loses its force. (ii) The frequentist interpretation leads to the locality condition used by John Bell in 1964 to establish conditions for the existence of hidden variables behind the spin probability distribution. It is shown that this locality condition is not consistent with the product (Bayes) rule of probability theory. Under the Bayesian interpretation, there is no motivation for this locality condition, which fails to preserve cogent information coming from the correct probabilistic conditioning. Indeed, a new stochastic hidden-variable model is given that reproduces the QM spin distribution, although it is not claimed that this model actually represents the underlying physics. (iii) As noted by some others and shown here using a simple proof that is independent of Bell's locality condition, Bell's original 1964 inequality involving expectations of pairs of spin variables does have an important role: it provides a necessary condition for the existence of a third-order joint probability distribution for three spin variables that is compatible through marginalization with the second-order joint distributions for the three possible spin pairs. An explicit expression is given here for this third-order distribution. Bell's original inequality is also shown to be logically equivalent to the 1969 CHSH Bell inequality that involves expectations of pairs of four spin variables; this inequality must be satisfied in order for a fourth-order joint probability distribution for the four spin variables to exist. When any of these Bell inequalities are violated, a joint probability distribution for three spin outcomes fails to be valid because some of the probabilities are negative.

**Keywords**: Bohm-EPR spin experiment, hidden-variable model, Bell inequalities, quantum entanglement, quantum non-locality, factorizability, Bayesian probability, frequentist probability

# 1. INTRODUCTION

QM (quantum mechanics) is a probabilistic mechanics theory that is unlike other such theories in that probability was not explicitly built into its foundations. Soon after Erwin Schrödinger's original publication (1926) in which the mysterious wave function emerged, Born (1926) postulated that the amplitude squared of the wave function for a particle should be viewed as a probability density function for the position of the particle. Later, Born's postulate was put on a more rigorous mathematical foundation by von Neumann (1932) and Gleason (1957) as what is now called Born's rule, which states that probabilities are given by certain inner products involving projections that act on the wave functions representing states of a system. Of course, for a quantity to be called probability it is essential that it satisfy the axioms of probability theory. It has not been unambiguously demonstrated that Born's rule does indeed imply satisfaction of all of the axioms. Ballentine (1986) addresses this issue. He argues that it is so, except that the product rule (Bayes rule) may not hold when it is applied to the values of incompatible variables whose corresponding operators do not commute. Goyal and Knuth (2011) explicitly *derive* Feynman's rules (Feynman 1948) for path probabilities in QM and show that the rules are consistent with probability theory if an assumption from classical physics is dropped. For this paper, the widely-held assumption that Born's rule does indeed deliver probabilities is accepted and the interest lies in how these probabilities should be interpreted.

Although there is widespread agreement on the axioms of probability, the meaning of probability has been debated for more than two centuries. However, little attention was paid to this issue in the early history of QM because the dominant interpretation of probability at the time was the frequentist one, which views probability as a property of inherently random events that is exhibited through their relative frequency of occurrence. More recently, the original Bayesian interpretation of probability (Bayes 1763; Laplace 1774, 1812), which quantifies a degree of belief or plausibility of a proposition conditional on specified information, has seen a renaissance throughout science and engineering because of its generality as a means of quantifying uncertainty when the available information is insufficient to make precise predictions. This renaissance has not yet had much impact on the interpretation of QM, although two physicists, Richard T. Cox and Edwin T. Jaynes, have made profound contributions to the rigorous foundations of Bayesian probability as a multi-valued logic for quantitative plausible reasoning (Cox 1946, 1961; Jaynes 1983, 2003). As a point of interest, two papers on the foundations of probability theory by Schrödinger show that he favored what would now be called a Bayesian interpretation of probability (Schrödinger 1947a,b).

The purpose of this work is to examine the implications of the frequentist and Bayesian meanings of probability on the interpretation of QM. This goal addresses one of the Oxford Questions discussed at the 2010 conference on Quantum Physics and the Nature of Reality: "How do different concepts of probability contribute to interpreting quantum theory?" (Briggs et al. 2013). The examination is done by using the Bohm-EPR experimental setup as a testbed. This setup is paradigmatic for some of the apparent mysteries of QM and it has been extensively studied both theoretically and experimentally, with pioneering contributions made by Bell (1964, 1987). However, most of these studies have implicitly taken a frequentist, rather than a Bayesian, interpretation of the joint probability distribution for the particle spins in these experiments.

In Section 2, the Bohm-EPR experimental setup is briefly presented to allow the paper to be self-contained and accessible to a broader audience interested in the meaning of probability. The focus then shifts to describing the frequentist and Bayesian interpretations of probability, and discussing their very different conclusions regarding the important issue of quantum non-locality. In Section 3, the differing implications of the two meanings of probability on the possible existence of hidden variables in Bohm-EPR experiments are investigated, along with the relevance of Bell inequalities. Bell's locality condition is motivated by a frequentist perspective but its special factorizing of the joint probability of the spin outcomes is shown to violate the product rule of probability theory. A stochastic hidden variable model is then presented that reproduces the QM spin distribution and is consistent with the probability axioms. Finally in Section 3, the relevance of Bell inequalities to an interesting problem in probability theory is discussed, along with its implications for the QM spin probabilities. Section 4 gives concluding remarks.

## 2. BOHM-EPR EXPERIMENTS AND THE ISSUE OF QUANTUM NON-LOCALITY

**Quantum mechanics probability distribution for the singlet state of two coupled spins**

Consider the gedanken experiment that was proposed by Bohm (1951, pp. 614-619) as an alternative way to frame the argument put forth by Einstein, Podolsky and Rosen (1935) that QM as it stands is an incomplete theory: there is a source that generates a particle that has a net spin of zero but immediately splits into two spin-½ charged particles, labeled A and B, that freely move apart in opposite directions. They each eventually enter a Stern-Gerlach (SG) device, with corresponding labels A and B, whose longitudinal axes lie along the line of motion of the particles. These devices can be rotated about their axes but their orientation is assumed known for each test. The "up" direction of devices A and B are denoted by unit directional vectors **a** and **b**, respectively, and they lie in a plane orthogonal to the longitudinal axes of the devices. The outcome of each SG device is unpredictable but binary: the particle is either deflected up or down. The corresponding outcomes are denoted by the binary variables A(**a**) = 1 (*spin up*) or $-1$ (*spin down*), and B(**b**) = 1 or $-1$, that is, "A(**a**)=1", for example, denotes the proposition, or event, that for a particle entering SG device A oriented in direction **a**, the spin outcome is up. There is no widely accepted explanation for the indeterminacy in these spin outcomes. From a frequentist perspective, it is assumed to be due to *inherent randomness*. From a Bayesian perspective, it is assumed to be due to *missing information*. The implications of these differing perspectives are discussed in later sub-sections.

QM theory, based on applying Born's rule to a spin wave function $\psi_{ss}$ for the *singlet state*, provides a joint probability distribution over the four possible pairs of spin values given by (A(**a**), B(**b**)) = $(1,1), (1,-1), (-1,1)$ and $(-1,-1)$ (Sakurai 2011):

$$
\begin{aligned}
P[A(\mathbf{a}), B(\mathbf{b})] &= \tfrac{1}{4}[1 - A(\mathbf{a})B(\mathbf{b})\mathbf{a} \cdot \mathbf{b}] \\
&= \tfrac{1}{2}\sin^2(\theta_{ab}/2) \quad \text{if } A(\mathbf{a}) = B(\mathbf{b}), \\
&\text{or } \tfrac{1}{2}\cos^2(\theta_{ab}/2) \quad \text{if } A(\mathbf{a}) = -B(\mathbf{b})
\end{aligned}
\qquad (1)
$$

where $\theta_{ab}$ is the angle between the unit vectors **a** and **b**. Although $\mathbb{P}[\ldots]$ will be used to denote the probability of an event or proposition, it is convenient to use a shorter notation for probability distributions where P[A(**a**),B(**b**)] in (1) denotes a probability function that gives the probability that (A(**a**), B(**b**)) equals a binary pair from the set $\{-1,1\} \times \{-1,1\}$; e.g. P[$\alpha,\beta$] = $\mathbb{P}$[A(**a**)=$\alpha$ and B(**b**)=$\beta$ | **a**,**b**,$\psi_{ss}$], where the conditioning indicates that **a** and **b** are assumed given and the two particles are in the singlet state defined by the wave function $\psi_{ss}$. This conditioning is left as understood in the shorter notation. Similarly, P[A(**a**)|B(**b**)] denotes a conditional probability $\mathbb{P}$[A(**a**)=$\alpha$|B(**b**)=$\beta$,**a**,**b**,$\psi_{ss}$] when the appropriate values ($\alpha,\beta$) of the two spin variables are substituted for (A(**a**), B(**b**)).

The marginal and conditional distributions corresponding to (1) may then be deduced for A(**a**) and B(**b**) (and, conversely, taking their product implies the joint distribution in (1)):

$$P[A(\mathbf{a})] = \sum_{B(\mathbf{b})=-1}^{+1} P[A(\mathbf{a}),B(\mathbf{b})] = 1/2, \quad P[B(\mathbf{b})] = \sum_{A(\mathbf{a})=-1}^{+1} P[A(\mathbf{a}),B(\mathbf{b})] = 1/2 \tag{2}$$

$$P[A(\mathbf{a}) \mid B(\mathbf{b})] = P[B(\mathbf{b}) \mid A(\mathbf{a})] = P[A(\mathbf{a}), B(\mathbf{b})] / P[A(\mathbf{a})] = \tfrac{1}{2}(1 - A(\mathbf{a})B(\mathbf{b})\, \mathbf{a}\cdot\mathbf{b})$$
$$= \sin^2(\theta_{ab}/2) \text{ if } A(\mathbf{a}) = B(\mathbf{b}),$$
$$\text{or } \cos^2(\theta_{ab}/2) \text{ if } A(\mathbf{a}) = -B(\mathbf{b}) \tag{3}$$

The main focus of this work is on the two interpretations, along with their differing implications, of the probability distribution in (1), and so on the equivalent distributions in (2) and (3). The conditional probabilities in (3) directly imply a correlation between the outcomes at the two SG devices, which is also exhibited by the result that the covariance is given by:

$$<A(\mathbf{a})B(\mathbf{b})> \triangleq \mathbb{E}[A(\mathbf{a})B(\mathbf{b})] \triangleq \sum_{A(\mathbf{a})=-1}^{+1} \sum_{B(\mathbf{b})=-1}^{+1} A(\mathbf{a})B(\mathbf{b})P[A(\mathbf{a}),B(\mathbf{b})] = \sin^2(\theta_{ab}/2) - \cos^2(\theta_{ab}/2)$$
$$= -\cos\theta_{ab} = -\mathbf{a}\cdot\mathbf{b} \tag{4}$$

and so does not factorize into $<A(\mathbf{a})><B(\mathbf{b})>$ (= 0). The underlying reasons for this correlation between the spin outcomes have been widely discussed, often under the topics of quantum non-locality, quantum entanglement, Bell inequalities and hidden variables. As we will show, the meaning of this correlation depends critically on whether probability is interpreted using a frequentist or a Bayesian perspective.

**Using the frequentist interpretation of probabilities**

A fundamental aspect of the frequentist interpretation of probabilities, which is commonly applied to QM theory, is that probability distributions are viewed as real properties of inherently random processes. For a set of possible random events that can occur, the probability P[E] of an outcome being an event E in this set is defined to be the relative frequency of its occurrence in a sequence of $n$ independent trials, repeated under similar conditions, as $n$ increases indefinitely. This definition of probabilities looks straightforward but its apparent simplicity cloaks some underlying difficulties.

In practice, it is often not possible to repeatedly perform a large number of independent trials and certainly not forever. The impracticality is even worse for continuous variables because they have an uncountable number of possible values; any attempt to directly establish to high precision the multi-dimensional probability distribution for a higher-dimensional vector of continuous variables would require a prohibitively large number of repeated trials. The frequentist definition is therefore conceptual rather than practical, so let us put practical difficulties aside and focus on the theoretical soundness of the definition.

The first point to notice is that it depends on an apparent limit as $n \to \infty$ but this mathematical limit cannot be directly defined because that requires examining the limiting behavior of an expression for the $n^{th}$ outcome, which does not exist for a random sequence of outcomes. The motivation for the definition is the law of large numbers (De Groot 1975) but this requires a prior definition of probability and so it cannot be used to define probability.

A second point is that the definition is based on an assumption that the same underlying conditions hold from trial to trial that somehow govern the randomness but it is not possible to define exactly what the required conditions are that need to be maintained and how they would work to control the randomness. This difficulty also carries over to applying frequentist probability to a single event, which requires viewing it as a sample from an *ensemble*, a reference set of events that are of the *same nature*. This embedding of the single event in an ensemble is often not conceptually possible because of the unrepeatable nature of the event, but even if it is possible, the choice of the ensemble is usually non-unique due to different possible choices of the conditions that the ensemble events must satisfy. This can lead to different probability distributions applied to the single event, such as in the *stopping rule problem* for sequential testing (O'Hagan and Forster 2004, p.121).

The final, and perhaps most important, issue is that the definition rests on the metaphysical assumption of the existence of inherently random processes but their existence cannot be demonstrated. For a process that looks random, there is a fundamental difficulty in distinguishing whether it is due to inherent randomness (ontic or aleatory uncertainty) or due to epistemic uncertainty because of missing information. For example, the outcome of tossing a coin and letting it land on a table is often described as being an inherently random process. It does look random in the absence of any information but if the initial conditions were known precisely (i.e. initial location and orientation of the coin and its initial angular velocity and center of mass velocity), then the outcome could be predicted accurately by rigid-body dynamics (e.g. Keller 1986). The apparent ontic uncertainty is actually epistemic uncertainty.

The Bohm-EPR experiment is often said to demonstrate through the experimental violation of Bell inequalities that there are no hidden variables and so the spin outcomes must be due to inherent randomness. This conclusion is doubtful, however, because the connection between Bell inequalities and the existence of hidden variables in the Bohm-EPR experiment is based on questionable arguments, which will be elaborated on in Section 3.

Applying the frequentist interpretation to the QM probabilities in (1)-(3), a conditional probability such as P[B(**b**)|A(**a**)] is viewed as an inherent physical property of the experimental setup. It is a logical inference from this premise that a measurement on the spin of one particle has a causal effect on the other particle. Thus, if A(**a**) is observed first, then it is implied that this outcome at SG device A must immediately establish the probability distribution P[B(**b**)|A(**a**)] for the outcome B(**b**) at SG device B; for the special case **b**=**a**, this implies that observing A(**a**)=1 must immediately cause B(**a**)=−1 (see (3)). The frequentist viewpoint therefore leads to the conclusion that there is a non-locality in Nature where an observation made by SG device A immediately affects the observation that will be made at SG device B. This apparent effect is sometimes referred to as a "superluminal quantum effect" (faster than the speed of light), while a skeptical Einstein called apparent behavior like this "spooky action at a distance" (Born 1971). Brunner et al. (2014) give a review of quantum non-locality with an extensive list of references. No plausible mechanism has been produced for an instantaneous influence between two particles at an arbitrary distance apart, nor is there any direct experimental evidence of it. The implication of this non-locality by the frequentist interpretation of probability therefore raises doubts about the validity of this interpretation. Other difficulties in applying frequentist probability have been widely discussed by statisticians (e.g. O'Hagan and Forster 2004), as well as physicists (e.g. Jaynes 1984; Loredo 1990).

**Using the Bayesian interpretation of probabilities**

The Bayesian interpretation of probability has a long history, starting with the paper by Thomas Bayes published posthumously in 1763 and the fundamental contributions of Pierre Simon Laplace, commencing in 1774 and culminating in his 1812 treatise (Laplace 1774, 1812). Because of its rigorous foundations, we use a specific Bayesian interpretation of probability that is primarily due to the seminal work of the physicists Richard T. Cox and Edwin T. Jaynes in which probability is viewed as providing an extended logic for the quantification of plausible reasoning under incomplete information (Cox 1946, 1961; Jaynes 1983, 2003). The probability $\mathbb{P}[b|c]$ is interpreted as the *degree of plausibility* of the proposition (statement) *b* based on the information stated in the proposition *c*, where *c* is only conditionally asserted but it cannot be self-contradictory (that is, it implies that a proposition is both true and false, and so is itself inherently false). Probability as a logic *extends* binary Boolean propositional logic, which is the special case where complete information is specified in a proposition *c* that gives the truth or falsity of *b*, that is, $\mathbb{P}[b|c]=1$ or $\mathbb{P}[b|c]=0$, respectively. For a quantitative propositional calculus for plausible reasoning, we need to evaluate the probabilities $\mathbb{P}[\sim b|c]$, $\mathbb{P}[a \& b|c]$ and $\mathbb{P}[a \text{ or } b|c]$ involving, respectively, a negation, a conjunction and a disjunction, in terms of more basic ones. These three probabilities correspond, respectively, to the degree of plausibility based on *c* that *b* is *not* true, that both *a* and *b* are true, and that either *a* or *b* (or both) are true. For this purpose, Cox (1946) postulated that universal *negation*, *conjunction* and *disjunction functions* of degrees of plausibility exist but he did not prescribe their explicit mathematical form. He then used the axioms of Boolean logic to *derive* the form of these universal functions. The result is that the calculus for treating uncertainty due to incomplete information is determined by the axioms for complete information to within an inconsequential smooth invertible mapping of the unit interval (Cox 1946; Jaynes 2003).

Cox's derived results can be stated as a minimal set of three axioms for probability as a logic (shortened here to probability logic):

For any propositions *a*, *b*, *c* with *c* not being self-contradictory:

(i) $\mathbb{P}[b|c] \geq 0$

(ii) $\mathbb{P}[\sim b|c] = 1 - \mathbb{P}[b|c]$

(iii) $\mathbb{P}[a \& b|c] = \mathbb{P}[a|b \& c]\mathbb{P}[b|c]$

Here, (ii) gives the negation function and (iii) gives the conjunction function (product rule or Bayes rule). For (iii) to be valid, propositions *b* and *c* cannot be contradictory because then *b*&*c* is a self-contradictory proposition and so $\mathbb{P}[a|b\&c]$ is undefined. The disjunction function (sum rule) can be derived from the last two axioms and De Morgan's Law from Boolean logic (Cox 1946; Jaynes 2003):

(iv) $\mathbb{P}[a \text{ or } b|c] = \mathbb{P}[a|c] + \mathbb{P}[b|c] - \mathbb{P}[a \& b|c]$

It is readily shown that these three axioms also imply that $\mathbb{P}[b|c] \in [0,1]$, which refines (i), and they lead to the well-known Marginalization, Total Probability and Bayes' Theorems found in any text book on probability theory (e.g. De Groot 1975). An overview of a framework for the application of probability logic to uncertainty quantification and propagation for dynamical systems is presented in Beck (2010) and Beck and Taflanidis (2013).

The axioms for a probability measure $\mathbb{P}(E)$ on subsets *E* of a finite set *X*, as stated by Kolmogorov (1950) and commonly given in textbooks on probability theory, can be derived as a special case of the probability logic axioms given above (Beck 2010). In Kolmogorov's formulation, the product rule comes from his definition of conditional probability as the ratio of two unconditional probabilities, a joint and a marginal, whereas in probability logic, all probabilities are conditional at the outset and the product rule is an axiom.

If $X$ denotes the set of possible values for an uncertain-valued variable $x$, then for any subset $E$ of $X$, the probability measure $\mathbb{P}(E)$ is interpreted in probability logic as $\mathbb{P}[x \in E | \pi]$ where $\pi$ denotes the proposition that states a probability model $f(x)$ quantifying the relative plausibility of each value of $x$ in $X$. The probability logic axioms therefore provide a calculus for handling *stochastic variables*, which is a terminology that is preferred for variables whose values are uncertain because of missing information, regardless of whether the variables correspond to physical quantities or to model parameters. The alternative terminology, *random variables*, which is common in frequentist probability theory, is not so appropriate in Bayesian probability theory because: (1) the fundamental concept is quantification of our uncertainty about both models and real events, and not describing Nature's apparent randomness; and (2) the definitions of stochastic and random variables differ from a technical point of view.

All probabilities interpreted from a Bayesian perspective are conditional on stated information, which at the very least must state, or logically imply, the probability models used to produce the values of the probabilities. Sometimes a probability is written in an unconditional form but this is only because the model being used for the probability distribution is not explicitly stated as conditioning information. Usually these models come from a theory, such as Born's rule in QM, but if not, a defensible way of choosing a model that is sometimes appropriate is to apply Jaynes' Principle of Maximum Entropy (Jaynes 1957) and maximize the Shannon entropy of the probability distribution subject to moment constraints. This principle avoids unjustified reductions in the uncertainty beyond that needed to enforce the constraints. From a Bayesian perspective, probabilistic conditioning is relevant information that informs the relative plausibility of the possible outcomes and it need not correspond to any direct causal connection with the outcomes. Jaynes (1989) gives a simple example of picking balls from an urn that nicely illustrates the distinction between causal dependence and information dependence.

Probability distributions for stochastic variables are viewed as chosen, or derived, models representing our uncertainty about their values and not as properties of Nature's "inherently random" events; indeed, the vague concept of inherent randomness, which needs to be postulated for the foundation of the frequentist approach, plays no role in probability logic. Instead, a pragmatic treatment of uncertainty for event outcomes is that the information needed to make perfect predictions is missing because of our limited capacity to access, or understand, the relevant information. In summary, Bayesian probability quantifies *our* uncertainty, and so it has an epistemic nature. It treats probability distributions as probability models for this purpose. On the other hand, frequentist probability is based on an ontological view where probability distributions are postulated to be real properties of inherently random events occurring in Nature, although in practice they are not known and must be modeled.

A difficulty that a newcomer may have when encountering the Bayesian definition of probability as a degree of conditional belief, or more specifically in our case, as a logic for quantitative plausible reasoning, is that it appears to have a more abstract nature than the frequentist definition. Then it may be helpful to point out the connection between probability and frequency of events that arises naturally in a Bayesian setting in the case where experiments or natural phenomena are of the nature of repeated events. It is based on the general procedure of using relevant data about stochastic variables to update their prior probability distribution through Bayes' Theorem. This connection can also be used to gain an appreciation, or "calibration", of a numerical probability value. Take a Bernoulli probability model for predicting whether or not an event occurs, then the most probable value *a posteriori* of the model parameter θ that gives the probability of this event occurring is equal to the relative frequency of its occurrence in any finite number of trials, when all values of θ are equally plausible *a priori*. The full Bayesian posterior distribution for θ is a Beta distribution over [0,1], which becomes more and more concentrated at the relative frequency that the event occurs as the number of trials increases (O'Hagan and Forster 2004, pp.4-5, 242). Bayes' Theorem can also be used to assess the probability of competing probability models based on relevant data; that is, one calculates the posterior probability for each model with respect to a set of proposed candidate models (Beck and Yuen 2004; Beck 2010).

The Bayesian interpretation of the probability distributions in (1)-(3) is that they are models derived from QM theory to make predictions about the uncertain experimental outcomes based on incomplete information, rather than being inherent physical properties of the experiment. The joint probability distribution in (1), and its equivalent in the form of (2) and (3), is implicitly conditional on the assumption that the two spin-½ particles are in the singlet state, since it is the wave function for this state that provides the probability distribution through an application of Born's rule.

A plausible explanation for the indeterminacy in the outcome of an SG device is that the spin direction of a spin-½ particle entering it is rotated by the magnetic field into an up or down direction with the outcome depending on the unknown initial spin direction and perhaps also on the spin phase (that is, where it is in its spin cycle). In this sense, rather than an SG device *measuring* the spin of the particle, it actually *transforms* the spin into either spin up or spin down, referred to as the *spin outcome* in this work. From this perspective, the probability distribution in (1) gives probabilistic predictions of the spin outcomes of the pair of particles upon exiting the SG devices, and not, as often stated, the probabilities for pre-existing states of spin up and spin down of the two particles entering SG devices A and B. For example, the conditional probability P[B(**b**)|A(**a**)] in (3) is viewed as a derived model that gives a measure of the plausibility of the outcome B(**b**) based on either actual or hypothesized information about the outcome A(**a**); recall that here A(**a**) and B(**b**) are shorthand for the statements that they are equal to the particular values that are to be substituted (e.g. B(**b**) represents either the statement B(**b**)=1 or B(**b**)=−1 ).

From the Bayesian perspective, conditioning in the probabilities for spin is viewed as providing information to be taken into account when making probabilistic predictions of the experimental spin outcomes, not as a causal influence of that outcome, so there is no implication that a measurement on particle A has any effect on particle B. This interpretation of conditional probability has also been advocated for QM by Jaynes (1989, 1990a,b), Fuchs (2003), De Raedt et al. (2007) and Grandy (2009).

The marginal distributions in (2) imply that all spin directions are equally plausible in the absence of any other information. The probability model in (1), which is conditional on the two-particle wave function for the singlet state, would not be appropriate if, for example, the direction of the coupled spins at the source was known. Actually, (1) corresponds to some ideal behavior for a real experiment because it depends only on a geometric projection and not on any details of an SG device.

Jaynes (1990a), Grandy (2009) and Goyal and Knuth (2011) have also advocated the Laplace-Cox-Jaynes approach to Bayesian probability as a probability logic for interpreting QM. Others have followed the alternative Ramsey-DeFinetti-Savage approach to subjective Bayesian probability (e.g. Fishburn 1986), which is based on betting odds, and is often referred to in the QM literature as Quantum Bayesianism or QBism (e.g. Caves, Fuchs and Shack 2002; Fuchs 2003; Pitowsky 2003; Fuchs, Mermin and Schack 2014).

**Global and local correlations: Bayesian information perspective**

*Equivalent predictive probability model for spin outcome distribution*

To gain further insight into the probability distribution (1) (and hence (2) and (3)), an equivalent predictive probability model for the two spin outcomes is presented that is a decomposition into global and local model properties. A Bayesian perspective is taken, so the probability distributions in this section are viewed as probability models that quantify the relative plausibility of the possible spin outcomes. In particular, we discuss the interpretation of the joint distribution P[A(**a**),A(**b**)] of the incompatible variables A(**a**) and A(**b**), which in the frequentist approach is not meaningful because both events cannot simultaneously occur and so this probability distribution cannot be a real property of the events.

*Equivalence Theorem*:
For all possible directions **a** and **b**, the joint probability distribution P[A(**a**),B(**b**)] in (1) over $\{-1,1\} \times \{-1,1\}$ and the corresponding probability distribution implied by the following two properties, are equivalent:

*Property I* (*Conservation of Total Spin*): For any direction **a**, A(**a**) = − B(**a**) = B(− **a**).

*Property II* (*Local Spin Correlation for a Stern-Gerlach Device*): For SG device A, the uncertainty in predicting the spin outcomes for the device for any two chosen directions **a** and **b,** is described by the joint probability distribution:

$$
\begin{aligned}
P[A(\mathbf{a}), A(\mathbf{b})] &= \tfrac{1}{4}[1 + A(\mathbf{a})A(\mathbf{b})\mathbf{a} \cdot \mathbf{b}] \\
&= \tfrac{1}{2}\cos^2(\theta_{ab}/2) \quad \text{if } A(\mathbf{a}) = A(\mathbf{b}), \\
&\text{or } \tfrac{1}{2}\sin^2(\theta_{ab}/2) \quad \text{if } A(\mathbf{a}) = -A(\mathbf{b})
\end{aligned} \quad (5)
$$

Property II is, in turn, equivalent to the conditional distribution:

$$
\begin{aligned}
P[A(\mathbf{b}) \mid A(\mathbf{a})] &= \tfrac{1}{2}[1 + A(\mathbf{a})A(\mathbf{b})\mathbf{a} \cdot \mathbf{b}] \\
&= \cos^2(\theta_{ab}/2) \quad \text{if } A(\mathbf{b}) = A(\mathbf{a}), \\
&\text{or } \sin^2(\theta_{ab}/2) \quad \text{if } A(\mathbf{b}) = -A(\mathbf{a})
\end{aligned} \quad (6)
$$

along with the marginal distributions: $P[A(\mathbf{a})] = \tfrac{1}{2} = P[A(\mathbf{b})]$ (this follows from the same approach as showing that (2) and (3) are equivalent to (1)).

*Proof*:

Property I is implied by the conditional probability distributions (3) from QM theory.

If **b** = **a** so that $\theta_{ab} = 0$, then from (3):
P[B(**a**) | A(**a**)] = 0  if A(**a**) = B(**a**),
            or  1  if A(**a**) = $-$ B(**a**)
so for sure, A(**a**) = $-$ B(**a**).

If **b** = $-$ **a** so that $\theta_{ab} = \pi$, then from (3):
P[B($-$**a**) | A(**a**)] = 1  if A(**a**) = B($-$**a**),
            or  0  if A(**a**) = $-$ B($-$**a**)
so for sure, A(**a**) = B($-$**a**). Then Property II follows by replacing B(**b**) in (1) by $-$ A(**b**).
Conversely, we can simply substitute A(**b**) = $-$ B(**b**) from Property I into Property II to get (1).

*Interpretation of Properties I and II*

Because of the Equivalence Theorem, a Bayesian interpretation of the QM probability distribution in (1) can be given in terms of Properties I and II.

An interpretation of Property I is that each particle has the physical property of a spin direction and the spins are created in opposite directions at their source, then, because of total spin conservation, the outcomes at two aligned SG devices are the opposite (in the absence of intervening electromagnetic-fields). Therefore, if SG device A produces A(**a**) = 1, then SG device B must produce B(**a**) = $-1$, while A(**a**) = $-1$ implies that B(**a**) = 1. This prediction based on the outcome at A can be made regardless of whether the companion particle has arrived at B or not; that is, the outcome at A provides information implying what the spin outcome at B will be, or actually is, regardless of the sequential order that the experimental outcomes occur. The model states, in effect, that when the two SG devices are aligned, it is spin conservation (total spin of zero) that provides the relevant information to infer the *global correlation* exhibited by the joint probability distribution in (1) for the predicted spin outcomes for particles A and B. This property is therefore the source of the quantum entanglement between the pair of spins.

Property II accounts for the partial mutual information between the spin outcomes from a single SG device. Although the two spin variables involved cannot be simultaneously observed for a particle, Property II can be readily interpreted. Consider first the conditional distribution in (6). Under the hypothesis that a particle entering SG device A along its longitudinal axis would produce *spin up* if the SG device was oriented in direction **a**, so that A(**a**) = 1, the probability that it would produce *spin up* for a new direction **b**, which is at an angle of rotation $\theta_{ab}$ about its longitudinal axis from direction **a**, is $\cos^2(\theta_{ab}/2)$. Under the same hypothesis, the probability that it would produce *spin down* for direction **b**, is $1 - \cos^2(\theta_{ab}/2) = \sin^2(\theta_{ab}/2)$. Similarly, under the hypothesis that A(**a**) = $-1$, the probabilities of the outcomes of spin down and spin up for direction **b** are $\cos^2(\theta_{ab}/2)$ and $\sin^2(\theta_{ab}/2)$, respectively. The value of the spin variable A(**a**) in the conditioning in (6) may be hypothesized or inferred to provide partial information for predicting the spin outcome A(**b**). If, for example, B(**a**) = $-1$ is the outcome at SG device B, then Property I implies that A(**a**) = 1. The conditional distribution in (6) therefore reveals a *local correlation* between the spin outcomes for any two directions of an SG device, which means that the specification of one of the spin outcomes provides information about what the other spin outcome would be. It is a local correlation because for Property II, the two directions **a** and **b** are for the same SG device.

As the direction **b** is rotated away from that of **a** through an angle $\theta_{ab}$, the given information about the outcome A(**a**) in (6) initially becomes of decreasing relevance and the uncertainty in A(**b**) increases. This increase is reflected in the conditional Shannon entropy for A(**b**) (Cover and Thomas 1991), which is a measure of uncertainty in the value of A(**b**), given A(**a**), due to missing information. This entropy increases from zero at $\theta_{ab} = 0$ to 1 bit at $\theta_{ab} = \pi/2$ where the probabilities for A(**b**) = 1 and A(**b**) = $-1$, conditional on A(**a**), are both equal to 1/2. Then the uncertainty, and hence the entropy of A(**b**), begins to decrease with further rotation until it is zero when **b** = $-$**a**, where the rotation angle is $\theta_{ab} = \pi$ and Property II implies the definite outcome that A(**b**) = $-$ A(**a**), that is, for sure A($-$**a**) = $-$ A(**a**).

The marginal probability distributions for A(**a**) and A(**b**) implied by Property II state that in the absence of any information about the spins, spin up and spin down are equally plausible, that is, for any direction **a**, $P[A(\mathbf{a})] = \frac{1}{2}$ for A(**a**) = 1 or $-1$. This distribution has a Shannon entropy of 1 bit, the largest entropy possible for a binary variable. Any other distribution for P[A(**a**)] would give smaller entropy, corresponding to a reduction in uncertainty that is not supported by any corresponding gain in information.

The local correlation specified by (5) can also be quantified by the mutual information I between A(**a**) and A(**b**) (Cover and Thomas 1991):

$$I[A(\mathbf{a}), A(\mathbf{b})] = \sum_{A(\mathbf{a})=-1}^{+1} \sum_{A(\mathbf{b})=-1}^{+1} P[A(\mathbf{a}), A(\mathbf{b})] \log_2 \frac{P[A(\mathbf{a}), A(\mathbf{b})]}{P[A(\mathbf{a})]P[A(\mathbf{b})]} = \tfrac{1}{2}\log_2(1-x^2) + \tfrac{1}{2}x\log_2\frac{(1-x)}{(1+x)} \quad (7)$$

where scalar $x = \mathbf{a} \cdot \mathbf{b}$. The mutual information I(*x*) is an even function of *x* and it is always positive except when **a** and **b** are orthogonal where I(*x*) has its minimum value of zero. It reaches a maximum value of 1 bit when **b** = **a** or **b** = $-$ **a**, which are the two cases where the value of A(**a**) implies the value of A(**b**), and conversely.

We conclude that, from a Bayesian perspective, the combination of the global and local correlations from Properties I and II, respectively, explains the relevance of the information from the spin outcome at one SG device to the probabilistic prediction of the outcome at the other device, without invoking any postulate about non-locality (instantaneous influence at a distance).

## 3. BOHM-EPR EXPERIMENTS: HIDDEN VARIABLES AND BELL INEQUALITIES

**Under the frequentist interpretation of probabilities**

The question of whether a hidden-variable model can explain the joint probability distribution in (1) for the pair of spin outcomes has a long history in which a prominent role is played by Bell inequalities, so named because the first such inequality was published by John Bell (1964). Bell inequalities are usually expressed in terms of covariances on three or four pairs of spin outcomes for Bohm-EPR experiments. They were originally derived under the assumption of hidden variables behind the indeterminacy of the spin outcomes in these experiments where the uncertainty in the values of the hidden variables is represented by a probability distribution. A specific locality assumption made by Bell is critical to these derivations. The inequalities are not satisfied by the covariances from QM that are given in (4) under some choices of the orientations of the two SG devices. The experimental evidence based on using *sample* covariances from multiple experiments to approximate the theoretical covariances in the Bell inequalities is usually taken as implying that the inequalities are violated. This has produced a commonly-held belief that there can be no hidden variables behind the QM results in (1), so we briefly review this argument.

*Bell's locality condition and implied covariance inequalities*

Bell (1964, 1981) assumed that if hidden variables (that is, properties associated with the particles) can explain the correlations exhibited in (1), then the following *locality condition* (also called the factorizability or factorability condition) must apply:

$$P[A(\mathbf{a}), B(\mathbf{b})| \lambda,\mathbf{a},\mathbf{b}] = P[A(\mathbf{a})| \lambda,\mathbf{a}] \, P[B(\mathbf{b})| \lambda,\mathbf{b}] \tag{8}$$

where $\lambda$ denotes the hypothesized hidden variables for the two "coupled" particles ("hidden" because they are not known explicitly). This factorization implies that $A(\mathbf{a})$ and $B(\mathbf{b})$ are independent conditional on $\lambda$ and so the conditional mean of the product is the product of the conditional means:

$$<A(\mathbf{a})B(\mathbf{b})| \lambda,\mathbf{a},\mathbf{b}> = <A(\mathbf{a})| \lambda,\mathbf{a}> <B(\mathbf{b})| \lambda,\mathbf{b}> \tag{9}$$

Bell's justification of his locality condition is that the orientation and outcome of SG device B should have no causal influence over the outcome of SG device A because of their physical separation. His argument is consistent with the frequentist interpretation of probability where conditional distributions are viewed as real properties of the experimental setup and the conditioning is viewed as a causal influence on the spin outcomes. The independences of $P[A(\mathbf{a})| \lambda,\mathbf{a}]$ from $\mathbf{b}$ and from $B(\mathbf{b})$ that are assumed in (8) are often called parameter and outcome independence, respectively (Shimony 1990).

Bell (1964) actually states his locality condition as a deterministic version of (8) that he calls the "vital assumption" that the outcome $B(\mathbf{b})$ for the particle at SG device B should not depend on the setting $\mathbf{a}$ for SG device B and uses the notation $A(\mathbf{a},\lambda)$ and $B(\mathbf{b},\lambda)$. He also imposes the condition $A(\mathbf{a},\lambda) = -B(\mathbf{a},\lambda)$, as in Property I. Later in Bell (1981), he again states that the outcomes at the two separated SG devices should not have any causal influence over each other but now expresses this fact in the probabilistic form in (8), presumably so that the outcomes at SG devices A and B need not necessarily be known when the orientations $\mathbf{a}$ and $\mathbf{b}$ and the hidden variable $\lambda$ are known, as he appears to assume in his 1964 paper. Bell showed that his locality condition leads to the following inequality in terms of three covariances:

$$|<A(\mathbf{a})B(\mathbf{b})> - <A(\mathbf{a})B(\mathbf{c})>| - <A(\mathbf{b})B(\mathbf{c})> \leq 1 \tag{10}$$

where $\mathbf{a}, \mathbf{b}$ and $\mathbf{c}$ are directions chosen for SG devices A and B and there is a total of three different Bohm-EPR experiments.

Clauser, Horne, Shimony and Holt (1969) derived another inequality using a similar approach to that in Bell (1964). The CHSH inequality involves four directions $\mathbf{a}$, $\mathbf{b}$, $\mathbf{c}$ and $\mathbf{d}$ for SG devices A and B and a total of four Bohm-EPR experiments:

$$|<A(\mathbf{a})B(\mathbf{b})> - <A(\mathbf{a})B(\mathbf{c})> + <A(\mathbf{d})B(\mathbf{b})> + <A(\mathbf{d})B(\mathbf{c})>|$$
$$\leq |<A(\mathbf{a})B(\mathbf{b})> - <A(\mathbf{a})B(\mathbf{c})>| + |<A(\mathbf{d})B(\mathbf{b})> + <A(\mathbf{d})B(\mathbf{c})>| \leq 2 \tag{11}$$

The first inequality in (11) is obvious. A simple proof of the second one based on (8) is given in Goldstein et al. (2011, p.8).

We show now that the Bell and CHSH inequalities in (10) and (11) are actually logically equivalent if, following Bell (1964), we assume that Property I holds. Set **d** = **b** in (11) so that <A(**d**)B(**b**)> = −1 because then A(**d**)=A(**b**)=−B(**b**) from Property I, implying that the final term in (11) is:
|<A(**d**)B(**b**)> + <A(**d**)B(**c**)>| = 1 − <A(**b**)B(**c**)>,
giving (10). Conversely, if we sum two Bell inequalities, one given by **a**, **b** and **c** as in (10), and the other by replacing them in order with **d**, **b** and −**c** in (10), then using B(−**c**)=−B(**c**) from Property I, we get the CHSH inequality in (11).

It is well known since their introduction that the Bell inequalities in (10) and (11) are violated for some choices of the directions by the QM distribution in (1). To first show this for (10), substitute the QM covariance expression in (4) into each term, then (10) can be expressed as:

$$|\cos\theta_{ab} - \cos\theta_{ac}| + \cos\theta_{bc} \leq 1$$

By taking $\theta_{ab} = \theta_{bc} = \pi/4$, so that $\theta_{ac}=\pi/2$, the left-hand side is $\sqrt{2}$ and the inequality is violated. Similarly, substituting the QM covariance expression in (4) into each term in (11), the CHSH inequality can be expressed as:

$$|\cos\theta_{ab} - \cos\theta_{ac} + \cos\theta_{db} + \cos\theta_{dc}| \leq 2$$

then for $\theta_{ab} = \theta_{db} = \theta_{dc} = \pi/4$, so that $\theta_{ac}=3\pi/4$, the left-hand side is $2\sqrt{2}$ and the inequality is violated. Actually, it is readily argued using continuity that there is a continuum of allowable values for the angles for which both inequalities are not satisfied.

To examine whether the experimental data on spin outcomes implies that the Bell inequalities are violated for some device orientations, each theoretical covariance, which corresponds to a single experimental set-up, is estimated by using sample covariances over many experiments. These empirical estimates of the theoretical covariances are then substituted into the CHSH inequality in (11). Most of these experiments have been performed using polarization of photons rather than spin-½ particles (e.g. Aspect et al. 1981, 1982; Shalm et al. 2015), so the experiments involve the spin-1 version of (1) where ½$\theta_{ab}$ is replaced by $\theta_{ab}$, but recently electron spins have been used (Hensen et al. 2015). The combined experimental evidence is taken to suggest that the CHSH inequality for photons and electrons is violated for certain choices of the orientations of the measuring devices, consistent with QM. It is then concluded that hidden variables that would explain the probability distribution (1) do not exist. This conclusion depends critically on the assumption that if hidden variables $\lambda$ exist, then the locality condition in (8) must hold, and the inference that the experimental results imply that (8) is violated for some choices of orientations.

*Bell's locality condition violates the product rule*

Actually, the factorizability (conditional independence) in (8) is fundamentally invalid because it is never possible for a hidden variable to provide all of the information that correct probabilistic conditioning provides, which is given by the product (Bayes) rule from probability theory:

$$P[A(\mathbf{a}),B(\mathbf{b})|\lambda,\mathbf{a},\mathbf{b}] = P[A(\mathbf{a})|B(\mathbf{b}),\lambda,\mathbf{a},\mathbf{b}]\, P[B(\mathbf{b})|\lambda,\mathbf{a},\mathbf{b}] = P[A(\mathbf{a})|\lambda,\mathbf{a},\mathbf{b}]\, P[B(\mathbf{b})|A(\mathbf{a}),\lambda,\mathbf{a},\mathbf{b}] \qquad (12)$$

For Bell's locality condition in (8) to be consistent with (12) requires that $P[A(\mathbf{a})|B(\mathbf{b}),\lambda,\mathbf{a},\mathbf{b}] = P[A(\mathbf{a})|\lambda,\mathbf{a}]$ but this equivalence is not compatible with Property I, the conservation of total spin, as shown next.

Consider the probability for $A(\mathbf{a})=1$ given that $B(\mathbf{b})=1$, and first set $\mathbf{b}=\mathbf{a}$, then from Property I, $A(\mathbf{a})=-B(\mathbf{a})=-B(\mathbf{b})=-1$, so $\mathbb{P}[A(\mathbf{a})=1|\, B(\mathbf{b})=1,\lambda,\mathbf{a}=\mathbf{b}] = 0$, regardless of the value of $\lambda$, which is set only by the source of the particles and has nothing to do with the settings of the SG devices. If instead $\mathbf{b}=-\mathbf{a}$ is chosen, then $A(\mathbf{a})=-B(\mathbf{a})=B(-\mathbf{a})=B(\mathbf{b})=1$, so $\mathbb{P}[A(\mathbf{a})=1|\, B(\mathbf{b})=1,\lambda,\mathbf{a}=-\mathbf{b}] = 1$. Thus, representing the probability $\mathbb{P}[A(\mathbf{a})=1|\, B(\mathbf{b})=1,\lambda,\mathbf{a},\mathbf{b}]$ as $\mathbb{P}[A(\mathbf{a})=1|\, \lambda,\mathbf{a}]$ is not valid because its value depends on the direction $\mathbf{b}$, which provides relevant information for the probability of $A(\mathbf{a})=1$ and so it cannot be dropped from the conditioning. When discussing the Bohm-EPR experiment, Jaynes (1989) noted that (12) is the correct factorization of the joint distribution according to the product rule but the implied incompatibility with Bell's locality condition under Property I does not seem to have been previously noticed.

Property I is not needed to prove the CHSH inequality (e.g. Goldstein et al. 2011, p.8). If it is dropped, however, then Bell's class of hidden variable models is inconsistent with the QM distribution in (1) at the outset. Such a class of models should therefore be of little interest. The essential point, though, is that Bell's locality condition does not apply to any class of hidden variable models that satisfies both the product rule for probabilities and Property I, the conservation of total spin. This fact is independent of whether the frequentist or Bayesian interpretation of probability is chosen.

**Under the Bayesian interpretation of probabilities**

From the perspective of Bayesian probability, the conditioning in (12) is viewed as relevant information for probabilistic predictions of the spin outcomes, rather than a causal influence on these outcomes. Thus, there is no concern about either superluminal propagation of effects or the need to show that there can be no instantaneous signaling by well-separated operators each using an SG device in a Bohm-Bell experimental setup. The underlying physical reason for the global correlation is the conservation of spin for the two particles. Furthermore, since Bell's locality condition in (8) is not applicable, the issue of whether hidden variables can explain the probability distribution in (1) is not settled by the experimental violation of the CHSH inequality using sample covariances that are constructed from multiple tests. To show that it is doubtful that any theorem can be developed to rule out hidden variables, a simple hidden variable model is presented, although it is not claimed that this model actually represents the underlying physics.

*Stochastic hidden variable model*

A stochastic hidden-variable model consistent with the joint probability distribution in (1) can be constructed as follows. Since (1) is equivalent to Properties I and II, and Property I has already been given a hidden-variable interpretation in terms of the two particles being created with opposite spin directions that explain the global correlation, we focus on Property II. We wish to demonstrate that there could exist hidden variables that explain the distribution P[A(**a**),A(**b**)] in (5) for an SG device.

Introduce the binary stochastic variable C(**a**,**b**) = A(**a**)A(**b**) with a distribution motivated by (5):

$\mathbb{P}[C=1| \mathbf{a},\mathbf{b}] = \frac{1}{2}(1 + \mathbf{a}\cdot\mathbf{b})$ and $\mathbb{P}[C=-1| \mathbf{a},\mathbf{b}] = \frac{1}{2}(1 - \mathbf{a}\cdot\mathbf{b})$

or, in the simpler notation introduced in Section 1: $P[C| \mathbf{a},\mathbf{b}] = \frac{1}{2}(1 + C\mathbf{a}\cdot\mathbf{b})$.

Next, define the conditional joint distribution P[A(**a**),A(**b**)|C] as in Table 1. Notice that these probabilities do not depend explicitly on the device settings **a** and **b**. Also, the marginal distribution for P[A(**a**),A(**b**)|C] is P[A(**a**)|C] = $\frac{1}{2}$ and so from Table 1, the conditional distribution P[A(**b**)|A(**a**),C] is deterministic (either 0 or 1), as expected, since if both C = A(**a**)A(**b**) and A(**a**) are specified, then A(**b**) is known.

From the Total Probability Theorem and Table 1:

$$P[A(\mathbf{a}),A(\mathbf{b})| \mathbf{a},\mathbf{b}] = \sum_{C=-1}^{+1} P[A(\mathbf{a}),A(\mathbf{b})|C]\, P[C| \mathbf{a},\mathbf{b}]$$

$$= \tfrac{1}{4}(1 + \mathbf{a}\cdot\mathbf{b}) \quad \text{if } A(\mathbf{a}) = A(\mathbf{b}),$$

$$\text{or } \tfrac{1}{4}(1 - \mathbf{a}\cdot\mathbf{b}) \quad \text{if } A(\mathbf{a}) = -A(\mathbf{b}) \tag{13}$$

which is the probability distribution in (5).

Table 1. Probability distribution P[A(**a**),A(**b**)| C] where C = A(**a**)A(**b**).

| (A(**a**),A(**b**)) | (+1,+1) | (+1,−1) | (−1,+1) | (−1,−1) |
|---|---|---|---|---|
| C=+1 | $\frac{1}{2}$ | 0 | 0 | $\frac{1}{2}$ |
| C=−1 | 0 | $\frac{1}{2}$ | $\frac{1}{2}$ | 0 |

A hidden variable is now introduced for C as follows. Let **λ** be a unit vector defining a direction, or, equivalently, defining a point on the surface *S* of an imaginary sphere of unit radius moving with the particle at its center. Roughly speaking, **λ** is like a spin direction for the particle. Partition the total surface *S* into two simply connected regions, $S_+(\mathbf{a},\mathbf{b})$ and $S_-(\mathbf{a},\mathbf{b})$, of areas $2\pi(1+\mathbf{a}\cdot\mathbf{b})$ and $2\pi(1-\mathbf{a}\cdot\mathbf{b})$, respectively. Take the conditional probabilities:

$\mathbb{P}[C=1|\boldsymbol{\lambda},\mathbf{a},\mathbf{b}] = 1$ if $\boldsymbol{\lambda} \in S_+(\mathbf{a},\mathbf{b})$ and $\mathbb{P}[C=-1|\boldsymbol{\lambda},\mathbf{a},\mathbf{b}] = 1$ if $\boldsymbol{\lambda} \in S_-(\mathbf{a},\mathbf{b})$
$= 0$ if $\boldsymbol{\lambda} \in S_-(\mathbf{a},\mathbf{b})$ $= 0$ if $\boldsymbol{\lambda} \in S_+(\mathbf{a},\mathbf{b})$ (14)

Take the hidden variable **λ** as uniformly distributed over the surface *S* and denote the probability density function for this distribution by p(**λ**). It does not depend on the device settings **a** and **b**. It follows that

$\mathbb{P}[\lambda \in S_+(\mathbf{a},\mathbf{b})] = \frac{1}{2}(1 + \mathbf{a} \cdot \mathbf{b})$ (it is the ratio of the area of $S_+(\mathbf{a},\mathbf{b})$ to the total area $4\pi$ of $S$) and, similarly, $\mathbb{P}[\lambda \in S_-(\mathbf{a},\mathbf{b})] = \frac{1}{2}(1 - \mathbf{a} \cdot \mathbf{b})$, so

$$\mathbb{P}[C=1|\mathbf{a},\mathbf{b}] = \int \mathbb{P}[C=1|\lambda,\mathbf{a},\mathbf{b}] \, p(\lambda) \, d\lambda = \int_{S_+} p(\lambda) \, d\lambda = \mathbb{P}[\lambda \in S_+(\mathbf{a},\mathbf{b})] = \frac{1}{2}(1 + \mathbf{a} \cdot \mathbf{b})$$
$$\mathbb{P}[C=-1|\mathbf{a},\mathbf{b}] = \int \mathbb{P}[C=-1|\lambda,\mathbf{a},\mathbf{b}] \, p(\lambda) \, d\lambda = \int_{S_-} p(\lambda) \, d\lambda = \mathbb{P}[\lambda \in S_-(\mathbf{a},\mathbf{b})] = \frac{1}{2}(1 - \mathbf{a} \cdot \mathbf{b}) \qquad (15)$$

Substituting the results in (15) into (13) gives a stochastic hidden-variable model for (5) of the form:

$$P[A(\mathbf{a}),A(\mathbf{b})|\mathbf{a},\mathbf{b}] = \int P[A(\mathbf{a}),A(\mathbf{b})|\lambda,\mathbf{a},\mathbf{b}] \, p(\lambda) \, d\lambda \qquad (16a)$$

$$P[A(\mathbf{a}),A(\mathbf{b})|\lambda,\mathbf{a},\mathbf{b}] = \sum_{C=-1}^{+1} P[A(\mathbf{a}),A(\mathbf{b})|C] \, P[C|\lambda,\mathbf{a},\mathbf{b}] \qquad (16b)$$

Given $\lambda$, C is known from (14), so then the joint probability of $A(\mathbf{a})$ and $A(\mathbf{b})$ is known from Table 1.

By substituting $A(\mathbf{b})=-B(\mathbf{b})$ from Property I throughout this subsection, (15) also gives a stochastic hidden-variable model for (1).

**Significance of Bell's inequality in probability theory**

Given that Bell's locality condition is invalid, it may look as if the Bell inequalities are irrelevant. It is a remarkable fact, however, that the Bell inequalities in (10) and (11) have a fundamental role in probability theory that is unrelated to Bell's locality condition or hidden variables. As noted by Hess and Philipp (2005), it was shown in the mathematical literature before Bell's work (e.g. Bass 1955; Vorob'ev 1962) that the inequality in (10) is part of a necessary condition for the existence of a valid third-order probability distribution $P[A(\mathbf{a}), A(\mathbf{b}), B(\mathbf{c})]$ for the three binary stochastic variables $A(\mathbf{a})$, $A(\mathbf{b})=-B(\mathbf{b})$ and $B(\mathbf{c})$ that gives by marginalization three second-order probability distributions for the three possible pairs of these variables. Furthermore, Fine (1982) showed that the existence of a valid fourth-order probability distribution $P[A(\mathbf{a}), A(\mathbf{d}), B(\mathbf{b}), B(\mathbf{c})]$ directly implies the eight inequalities on the second-order joint probabilities that were first derived by Clauser and Horne (1974) based on Bell's locality condition. Furthermore, these eight probability inequalities can be expressed in terms of covariances, leading to two CHSH inequalities implied by (11) by removing the absolute values, as well as six others that are just permutations of the directions $\mathbf{a}$, $\mathbf{b}$, $\mathbf{c}$, and $\mathbf{d}$ in these two inequalities. The CHSH inequality in (11) is therefore a necessary condition for the existence of a valid fourth-order joint distribution for the four spin variables. If (11) is violated, a third-order joint distribution fails to exist because it has negative probabilities and so, therefore, the fourth-order joint does not exist as well (a valid fourth-order joint implies all four third-order joints exist from marginalizations over each of the four variables).

A simple proof of these results is given in the Appendix. By taking $A=A(\mathbf{a})$, $B=A(\mathbf{b})$ and $C=A(\mathbf{c})$ in (A6) and (A7), a necessary condition for the existence of a valid joint distribution $P[A(\mathbf{a}), A(\mathbf{b}), A(\mathbf{c})]$ that is compatible with three specified second-order joint distributions $P[A(\mathbf{a}),A(\mathbf{b})]$, $P[A(\mathbf{a}),A(\mathbf{c})]$, and $P[A(\mathbf{b}),A(\mathbf{c})]$ can be expressed in terms of the corresponding covariances as:

$$|<A(\mathbf{a})A(\mathbf{b})> + <A(\mathbf{a})A(\mathbf{c})>| - <A(\mathbf{b})A(\mathbf{c})> \leq 1 \qquad (17a)$$

|<A(**a**)A(**b**)> − <A(**a**)A(**c**)>| + <A(**b**)A(**c**)> ≤ 1                                                                                    (17b)

If at least one of the two Bell inequalities in (17) is not satisfied, then a valid third-order distribution for A(**a**), A(**b**) and A(**c**) does not exist because at least one of the joint probabilities is negative. Bell's inequality in (10) corresponds to using Property I to substitute A(**b**) =− B(**b**) into the first term of (17b) and A(**c**) =− B(**c**) in the second and third terms of (17b), so it is a necessary condition for a valid joint distribution P[A(**a**), B(**b**), B(**c**)] or P[A(**a**), A(**b**), B(**c**)].

If in addition to the above substitutions for A, B and C, we also set D=A(**d**) in (A10), then:

|<A(**a**)A(**b**) − <A(**a**)A(**c**)>| + |<A(**d**)A(**b**)> + <A(**d**)A(**c**)>| ≤ 2                                    (18)

is a necessary condition for a valid joint distribution P[A(**a**), A(**b**), A(**c**), A(**d**)]. If Property I is used to substitute A(**b**) =− B(**b**) and A(**c**) =− B(**c**), then (18) corresponds to the CHSH inequality (11). Therefore, if (11) is not satisfied, then the conclusion after (A10) implies that a valid fourth-order joint P[A(**a**), A(**d**), B(**b**), B(**c**)] does not exist because some of the probabilities are negative.

Under symmetry of the probability distributions with respect to flipping the spin labels {− 1,1} that are the binary values of A(**a**), A(**b**) and A(**c**), it is shown in the Appendix that (17a) and (17b) together give a *necessary and sufficient* condition for the existence of P[A(**a**), A(**b**), A(**c**)]. Furthermore, there is always a unique mathematical solution for the third-order distribution P[A(**a**),A(**b**),A(**c**)] that gives, by marginalization, three specified second-order distributions, P[A(**a**),A(**b**)], P[A(**b**),A(**c**)] and P[A(**a**),A(**c**)]. This third-order joint distribution is given by (A9) in the Appendix by substituting A=A(**a**), B=A(**b**) and C=A(**c**). In the case of QM, where the second-order joint distributions are given by (1), this compatible third-order distribution given by (A9) can be written as:

P[A(**a**),A(**b**),A(**c**)] = $\frac{1}{8}$[1 + A(**a**)A(**b**) **a** · **b** + A(**b**)A(**c**) **b** · **c** + A(**c**)A(**a**) **c** · **a**]                                   (19)

where the covariances in (A9) have been expressed by using (4), noting that B(**b**) =− A(**b**) by Property I.

In the case of the second-order joint distributions given by QM, we know that there exists coplanar directions **a**, **b** and **c** for which at least one of the inequalities (17a) or (17b) is not satisfied. For these directions, the implied third-order joint distributions are not valid because some of the probabilities are negative. A similar conclusion applies to the CHSH inequality (11), which under Property I effectively involves the four variables A(**a**), A(**b**), A(**c**) and A(**d**) and therefore requires a fourth-order joint probability distribution, P[A(**a**), A(**b**), A(**c**), A(**d**)]. If (11) is not satisfied, a third-order joint distribution fails to exist because it has negative probabilities and so, therefore, the fourth-order joint does not exist as well (a valid fourth-order joint implies all four third-order joints exist).

Only a pair of directions, say **a** and **b** for SG devices A and B, respectively, can be examined in any one experiment and QM theory delivers a joint probability distribution that applies to predicting the outcomes A(**a**) and B(**b**)=− A(**b**). The fact that the theory cannot provide third or higher joint distributions for predicting the spin outcomes for some choices of three or more different SG device orientations is therefore not a deficiency from the perspective of predicting outcomes in any one Bohm-EPR experiment. However, it is puzzling that for some orientations, QM delivers probabilistic predictions for A(**a**) and A(**b**)=− B(**b**) in the absence of any information (since P[A(**a**),A(**b**)] exists), but not under the information that A(**c**)=−B(**c**) is known (because P[A(**a**),A(**b**)|A(**c**)] = P[A(**a**),A(**b**),A(**c**)]/P[A(**c**)] fails to exist). For other choices of SG device orientations, these unconditional and conditional second-order joint distributions both exist, despite the fact that A(**a**) and A(**b**) are incompatible variables and their spin operators do not commute, so QM theory is not expected to deliver such joint probabilities.

## 4. CONCLUSIONS

The Bohm-EPR gedanken experiment serves as an important test bed to examine the differing implications of the two principal meanings of axiomatic probability, frequentist and Bayesian, on interpretations of QM theory. The focus is on the probability distribution derived from Born's rule for the spin outcomes of two particles with coupled spins in the singlet state that are produced by separated SG (Stern-Gerlach) devices.

It is argued that the widely-discussed apparent non-locality based on the implied probabilistic spin correlations from QM is a consequence of the frequentist interpretation of the probabilities: a probability distribution is viewed as a real property of "inherent randomness" and so conditioning in the probabilities for an event is viewed as playing a causal role for the event. Therefore, observing the spin outcome at one device immediately influences the spin outcome at the other device, although no plausible mechanism has been presented for such a superluminal effect. From a Bayesian perspective, the probabilistic spin correlations mean that the spin outcome (actual or hypothesized) at one SG device provides information relevant to the probability for predicting the spin outcome at the other device, and there is no motivation to postulate any non-local effect. It is argued that the predicted spin correlations have both a local and a global component and the source of the global correlations is the conservation of the total spin of the two spin-coupled particles, which is therefore also the source of the quantum entanglement.

For the question of whether a hidden-variable model can explain the joint probability distribution for the pair of spin outcomes that is given by QM theory, it is argued that Bell inequalities are irrelevant. Bell's locality (factorability/factorizability) condition that he and others have used to establish inequalities that must be satisfied for the existence of hidden variables behind the spin probability distribution is motivated by the frequentist interpretation of probability. However, it is inconsistent with the product rule of probability theory. This leaves as an open question the possible existence of hidden variables behind the QM spin distribution. Indeed, a stochastic hidden variable model is presented that reproduces this distribution and is consistent with the probability axioms, although it is not claimed that this model actually represents the underlying physics.

Bell inequalities do have an important role: they give a necessary condition for the existence of a third-order (or fourth-order) joint probability distribution for the spin outcomes that has as marginal distributions, three (or four) QM second-order joint probabilities. A simple proof of this result that does not use Bell's locality condition is given. For choices of the orientations of the SG devices where Bell inequalities are not satisfied, some of the third-order joint probabilities become negative. Repeated experiments with a pair of orientations (**a**,**b**) can be used to check the joint probability distribution for A(**a**) and B(**b**) from QM by using the sample moments to estimate the corresponding theoretical moments in equation (A8). However, performing a set of separate experiments for each of three or four pairs of orientations, (**a**,**b**), (**b**,**c**) and so on, and then substituting the sample moments into the Bell inequalities (10) or (11), cannot be used to check for the existence of hidden variables underlying the spin distribution.

In summary, the frequentist interpretation of probability treats probability distributions as real properties of random phenomenon that control the long-term behavior of apparently random events through some invisible hand. In QM, this interpretation leads to apparent quantum non-locality and to Bell's locality condition that violates the product rule of probability theory. From the perspective of E.T Jaynes, the frequentist interpretation of probability is an example of what he calls the Mind-Projection Fallacy where models of reality are confused with reality (Jaynes 1990a,b, 2003).

If the Bayesian interpretation of probability is chosen, there is no motivation for quantum non-locality or Bell's locality condition. Probability distributions are then viewed pragmatically as probability models for predicting outcomes that are uncertain because there is insufficient information available for precise predictions, and conditioning in probability distributions is viewed as information to be used in inferences and not as representing a causal influence. The adoption of the Bayesian point of view leads to less puzzling interpretations of QM theory than the frequentist perspective in understanding quantum entanglement in the Bohm-EPR experiment. It also has important implications for understanding other aspects of QM, such as the quantum measurement problem.

**APPENDIX: Bell inequalities and existence of third and fourth order distributions**

Let A, B and C be any three binary stochastic variables whose possible values are $\{-1,1\}$. If a third-order joint distribution exists for A, B and C, then it can be expressed in the form:

$$\mathbb{P}[A=\alpha, B=\beta, C=\delta] = \tfrac{1}{8}[1 + \alpha\langle A\rangle + \beta\langle B\rangle + \delta\langle C\rangle + \alpha\beta\langle AB\rangle + \beta\delta\langle BC\rangle + \alpha\delta\langle CA\rangle + \alpha\beta\delta\langle ABC\rangle]$$

where $\alpha, \beta, \delta = +1$ or $-1$, or, equivalently, using the shorthand notation for probability functions:

$$P[A,B,C] = \tfrac{1}{8}[1 + A\langle A\rangle + B\langle B\rangle + C\langle C\rangle + AB\langle AB\rangle + BC\langle BC\rangle + CA\langle CA\rangle + ABC\langle ABC\rangle] \qquad (A1)$$

This result can be shown by noting that it is the unique solution for the eight probabilities defining the distribution that satisfies normalization and the seven moment equations.

Suppose now that the values of the seven moments are specified in the interval $[-1,1]$ and we ask if (A1) gives a valid probability distribution. This depends on whether all eight probabilities lie in the interval $[0,1]$. Clearly, these probabilities are bounded above by 1 because the magnitude of each of the eight terms inside the bracket in (A1) is bounded above by 1. However, depending on the specified values of the moments, P[A,B,C] may be negative. Necessary and sufficient conditions for P[A,B,C] to be a valid probability distribution for specified values of the seven moments are the eight conditions on the moments implied by the right-hand side of (A1) being non-negative as A,B and C range over $-1$ and 1. These eight conditions come in four pairs where each pair corresponds to switching the sign of each component of the triplet (A,B,C), namely pairs $(1,-1,-1)$ and $(-1,1,1)$, $(1,1,1)$ and $(-1,-1,-1)$, $(1,-1,1)$ and $(-1,1,-1)$, and $(1,1,-1)$ and $(-1,-1,1)$. If we sum each such pair, we get four Bell inequalities involving only the second-order moments:

$$\langle AB\rangle + \langle BC\rangle + \langle AC\rangle \geq -1 \qquad (A2)$$

$$\langle AB\rangle - \langle BC\rangle - \langle AC\rangle \geq -1 \tag{A3}$$
$$-\langle AB\rangle + \langle BC\rangle - \langle AC\rangle \geq -1 \tag{A4}$$
$$-\langle AB\rangle - \langle BC\rangle + \langle AC\rangle \geq -1 \tag{A5}$$

These four inequalities can be reduced to two equivalent inequalities that each have a similar form to the original 1964 Bell inequality in (10) by combining (A2) and (A4), and (A3) and (A5):

$$|\langle AB\rangle + \langle AC\rangle| - \langle BC\rangle \leq 1 \tag{A6}$$
$$|\langle AB\rangle - \langle AC\rangle| + \langle BC\rangle \leq 1 \tag{A7}$$

Satisfaction of these two Bell inequalities is therefore necessary for a valid third-order distribution for (A,B,C). If either of them is violated, then a valid third-order distribution for (A,B,C) does not exist with the three specified second moments.

If the three second-order distributions P[A,B], P[B,C] and P[C,A] are given, which imply the marginal distributions P[A], P[B] and P[C], then it follows that the first six moments are specified because

$$P[A,B] = \tfrac{1}{4}[1 + A\langle A\rangle + B\langle B\rangle + AB\langle AB\rangle], \quad P[A] = \tfrac{1}{2}[1 + A\langle A\rangle], \text{ etc.} \tag{A8}$$

Then with an arbitrary choice of the third moment $\langle ABC\rangle$, the third-order distribution P[A,B,C] given by (A1) will be a valid one provided all eight probabilities are non-negative. The necessary conditions for this to hold are then satisfaction of (A6) and (A7), which can also be expressed in terms of probabilities by using (A8) to replace the moments.

In the special case where the joint probabilities for P[A,B,C] are symmetric with respect to the binary values of these variables, the odd moments are zero and (A1) reduces to

$$P[A,B,C] = \tfrac{1}{8}[1 + AB\langle AB\rangle + BC\langle BC\rangle + CA\langle CA\rangle] \tag{A9}$$

In this case, satisfaction of the two inequalities (A6) and (A7) gives a *necessary and sufficient* condition for a valid third-order distribution. This proves the following theorem:

*Existence Theorem*: *A necessary condition for a valid third-order probability distribution* P[A,B,C] *to exist that gives by marginalization three specified second-order distributions* P[A,B], P[A,C] *and* P[B,C] *is that the four inequalities (A2-A5) are all satisfied, or, equivalently, the two inequalities in (A6) and (A7) are satisfied. If the joint probabilities for P[A,B,C] are symmetric with respect to flipping the signs of all three variables, then satisfaction of the inequalities (A6) and (A7) gives a necessary and sufficient condition.*

If we have four binary stochastic variables A, B, C and D with a fourth-order distribution P[A,B,C,D], then by marginalization, the two third-order distributions for (A,B,C) and (B,C,D) exist. The Existence Theorem then implies that (A7) holds and so does (A6) with A replaced by D. Summing these two inequalities then implies an inequality of the form of the 1969 CHSH inequality in (11):

$$|\langle AB\rangle - \langle AC\rangle| + |\langle DB\rangle + \langle DC\rangle| \leq 2 \tag{A10}$$

This inequality is therefore a necessary condition for the existence of a fourth-order distribution for (A,B,C,D) that is compatible with the second-order distributions for the pairs (A,B), (A,C), (D,B) and (D,C). If inequality (A10) is not satisfied, then at least one of the joint probabilities for (A,B,C,D) is negative and so a valid fourth-order joint distribution P[A,B,C,D] does not exist. If it did, then all four third-order marginal distributions would exist, contradicting the assumed violation of inequality (A10). Three other similar but distinct necessary conditions can be obtained by examining permutations of A, B, C and D in (A10). Using (A8), these four inequalities can also be expressed in terms of second-order joint and first-order marginal probabilities to give the counterparts of the four inequalities in Clauser and Horne (1974) and Fine (1982).

Here is a simple example motivated by one in Vorob'ev (1962) that illustrates the Existence Theorem. Consider binary stochastic variables A, B and C that have second-order joint distributions as in Table 2.

Table 2. Probability distributions P[A,B], P[B,C] and P[C,A].

| P[A,B] | A=+1 | A=−1 | P[B,C] | B=+1 | B=−1 | P[C,A] | C=+1 | C=−1 |
|---|---|---|---|---|---|---|---|---|
| B=+1 | $\frac{1}{2}$ | 0 | C=+1 | 0 | $\frac{1}{2}$ | A=+1 | $\frac{1}{2}$ | 0 |
| B=−1 | 0 | $\frac{1}{2}$ | C=−1 | $\frac{1}{2}$ | 0 | A=−1 | 0 | $\frac{1}{2}$ |

The first and second moments of A, B and C are <A>=<B>=<C>=0 and <AB>=<CA>=1 and <BC>=−1. Substitution of these moments shows that inequality (A7) is satisfied but that inequality (A6) is violated. Therefore, a valid third-order distribution P[A,B,C] does not exist and there must be at least one negative probability. The mathematical expression for it in (A1) gives:

$$P[A,B,C] = \tfrac{1}{8}[1 + AB - BC + CA + ABC\mu_3] \tag{A11}$$

where $\mu_3$ = <ABC>. This third moment is bounded by 1 so the probability for A=−1 and B= C= 1 is negative.